\documentclass[9pt,twocolumn,a4paper]{article}

\usepackage[utf8]{inputenc}
\usepackage[margin=1.5cm]{geometry}
\usepackage{graphicx}
\usepackage{amsmath,amssymb}
\usepackage{authblk}
\usepackage{titlesec}
\usepackage{xcolor}
\usepackage{hyperref}
\usepackage{caption}
\usepackage[superscript,biblabel]{cite}

\definecolor{natureblue}{RGB}{21,31,211}
\hypersetup{colorlinks=true, linkcolor=natureblue, citecolor=natureblue, urlcolor=natureblue}

\titleformat{\section}{\large\bfseries}{\thesection}{1em}{}
\titleformat{\subsection}{\normalsize\bfseries}{\thesubsection}{1em}{}

\date{}
\begin{document}

\title{
\fontsize{15}{17}\selectfont
\textbf{Magnetic-configuration design for reliable Heisenberg exchange parameters}
}

\author[1]{Ben Li}
\author[1,*]{Stephan von Malottki}
\author[1,2,3]{Gian-Marco Rignanese}

\affil[1]{Institute of Condensed Matter and Nanosciences, Universit\'{e} catholique de Louvain, Chemin des \'{E}toiles 8, B-1348 Louvain-la-Neuve, Belgium}
\affil[2]{WEL Research Institute, avenue Pasteur 6, B-1300 Wavre, Belgium}
\affil[3]{School of Materials Science and Engineering, Northwestern Polytechnical University, No. 127 Youyi West Road, Xi’an 710072, Shaanxi, China}
\affil[*]{Corresponding author: malottki@outlook.com}

\maketitle

\textbf{\begin{abstract}
The determination of magnetic exchange interactions is essential for the quantitative description and predictive modeling of magnetic materials. In this work, we present a neighbor-shell-based screening method for selecting magnetic configurations suitable for extracting Heisenberg exchange parameters beyond nearest-neighbor from density functional theory (DFT) calculations. Using only the structure, the proposed approach identifies the linear independence of neighbor-shell contributions before any first-principles calculations instead of relying on trial-and-error generation of magnetic configurations. We apply the proposed approach to two representative classes of magnetic configurations: random spin states and spin spirals, and validate its predictions against direct DFT fitting for Fe and MnF$_2$. We show that only parameters obtained when all relevant neighbor-shell contributions are linearly independent remain transferable to other magnetic configurations. The proposed method provides practical guidance for selecting magnetic configurations for reliable exchange-parameter extraction and may also benefit other neighbor-shell-based models.
\end{abstract}}

\section{Introduction}\label{sec1}

The thermodynamic description of magnetic materials has long been a challenge for solid-state theorists. Thanks to the development of DFT and modern first-principles methods, the ground-state properties of magnetic systems can now be described with high accuracy\cite{Hoffmann2020, Sato2010}. Nevertheless, the prediction of magnetic properties remains computationally demanding in practice \cite{Kotykhov2023, Makkar2021, Sokolovskiy2023}. This is particularly true for noncollinear magnetism, for which achieving stable electronic convergence is often difficult and the use of large supercells can be computationally prohibitive \cite{Manz2011}. Developing approaches that identify suitable magnetic configurations while reducing the simulation cell size is therefore of great practical importance. Such strategies can substantially decrease the computational cost and may also enable the application of more refined first-principles methods.

A common route to thermodynamic magnetic properties is to construct an effective Heisenberg Hamiltonian from first-principles calculations \cite{Liechtenstein1984, Hellsvik2019, Szilva2023} and then use it to evaluate quantities such as transition temperatures \cite{Pajda2001}. There are two main methods to derive such spin models. The first strategy, referred to as the real-space approach \cite{Liechtenstein1987}, determines the exchange parameters $J_{ij}$ directly from the energy variation induced by constrained rotations of the spin-polarization axes at sites $i$ and $j$. Within the framework of the so-called magnetic force theorem, this variation in the total energy is approximated by the corresponding change in the one-electron energies \cite{Szilva2013, Solovyev2021, Oswald1985}, which simplifies the evaluation of the exchange interactions. In the second mapping method, one fits the total energies of a large number of magnetic configurations from electronic structure calculations onto the spin Hamiltonian to get the exchange parameters \cite{Sandratskii1991, Xiang2011, fmmode2020}. In this work, we focus on the latter strategy, which is broadly applicable to variational electronic-structure methods that provide total energies, including calculations based on collinear states, random magnetic configurations, and spin spirals \cite{Szilva2023}. Collinear states, in which spins are constrained to flip along a fixed axis, are often sufficient for Ising-like models \cite{Satya2020}. In contrast, the Heisenberg model treats magnetic moments as vector degrees of freedom that can rotate in three-dimensional space; therefore it requires magnetic configurations that contain noncollinear spin information \cite{Szilva2023, Jacobsson2022}. Random spin states and spin spirals provide two representative ways to offer such information, making them suitable for describing complex magnetic systems with competing and frustrated exchange interactions. Among these approaches, spin spirals are attractive because they can be computed in the primitive cell without spin–orbit coupling (SOC) by the generalized Bloch theorem (GBT) \cite{Kurz2004}.

Considerable effort has been devoted to improving the accuracy of the total energies of magnetic configurations used in the mapping procedures. Beyond conventional DFT, other advanced methods such as DFT+U, dynamical mean-field theory (DMFT), hybrid functionals and GW approximation have been successfully employed to improve the correspondence between calculated and experimental properties \cite{Anisimov1991, Kune2007, Franchini2005, Faleev2004}. These developments improve the quality of the input energy used for spin-model construction. However, the reliability of the extracted exchange interactions is determined not only by the total-energy accuracy, but also by the choice of magnetic configurations used in the procedure. Including too many unnecessary configurations can significantly increase the computational cost and may even reduce the stability of the extracted parameters. Therefore, identifying suitable magnetic configurations before performing DFT calculations is essential for efficiently obtaining reliable exchange parameters. Recent works have analyzed whether a given set of magnetic configurations contains sufficient independent information to determine the targeted interactions, using null-space analyses of the corresponding fitting matrices \cite{Jacobsson2013, Alaei2025, Rezaei2025}. These approaches provide useful mathematical evaluation for identifying exchange parameters that are not uniquely determined from the selected configurations. However, such analyses do not fully explain how this non-uniqueness emerges from the intrinsic structural and magnetic characteristics of the material. As a result, a systematic strategy for designing magnetic configurations to extract exchange interactions prior to first-principles calculations is still lacking.

In this work, we develop an efficient method to select the optimized magnetic configurations for determining specific Heisenberg exchange parameters by analyzing the linear independence of the corresponding neighbor-shell contributions. By incorporating structural periodicity and the relative phases between magnetic sublattices, the method identifies which exchange shells can be independently resolved for a given family of magnetic configurations without performing any DFT calculations. The model predictions are then validated by direct comparison with DFT calculations for two representative classes of magnetic configurations: random magnetic states and spin-spiral configurations. In addition, we identify suitable examples of magnetic configurations for fitting and demonstrate the transferability of the extracted exchange parameters for body-centered cubic (bcc) Fe and $\mathrm{MnF_2}$.

\section{Results}\label{sec2}

\subsection{Linear independence criterion}

The exchange interaction is the dominant term in the Heisenberg model. For classical spins, the corresponding effective Heisenberg Hamiltonian is given by \cite{Frota2000, Pajda2001}
\begin{align}
H_{\mathrm{eff}}
= -\frac{1}{2}\sum_{ij}^{N} J_{ij}\,
\mathbf{e}_i \cdot \mathbf{e}_j,
\label{eq:1}
\end{align}
where $\mathbf{e}_{i}$ and $\mathbf{e}_{j}$ are unit vectors along the directions of the local magnetic moments at sites $i$ and $j$, respectively, and $N$ is the total number of magnetic sites. The extraction of parameters of other interactions such as Dzyaloshinskii-Moriya interaction (DMI), magnetocrystalline anisotropies or higher-order exchange interactions (HOI) is not the subject of this paper. However, they certainly can be treated with an analogous approach to the one presented here. In the mapping method used to extract the exchange parameters $J_{ij}$, multiple magnetic configurations generated from first-principles calculations are employed to fit an effective spin model. A common assumption is that the contributions from different neighbor shells are linearly independent within the chosen set of configurations \cite{Williams2016, Zhu2020, Logemann2017}. However, in actual calculations, symmetry relations and periodic boundary conditions can generate correlations between these contributions, which in turn restrict the number of neighbor-shell interactions that can be extracted robustly. Different neighbor-shell contributions may carry dependent information within a given family of magnetic configurations, and this limitation is general to energy-mapping approaches based on finite sets of magnetic states. 

\begin{figure}[htbp]
	\centering
	\includegraphics[width=0.8\linewidth]{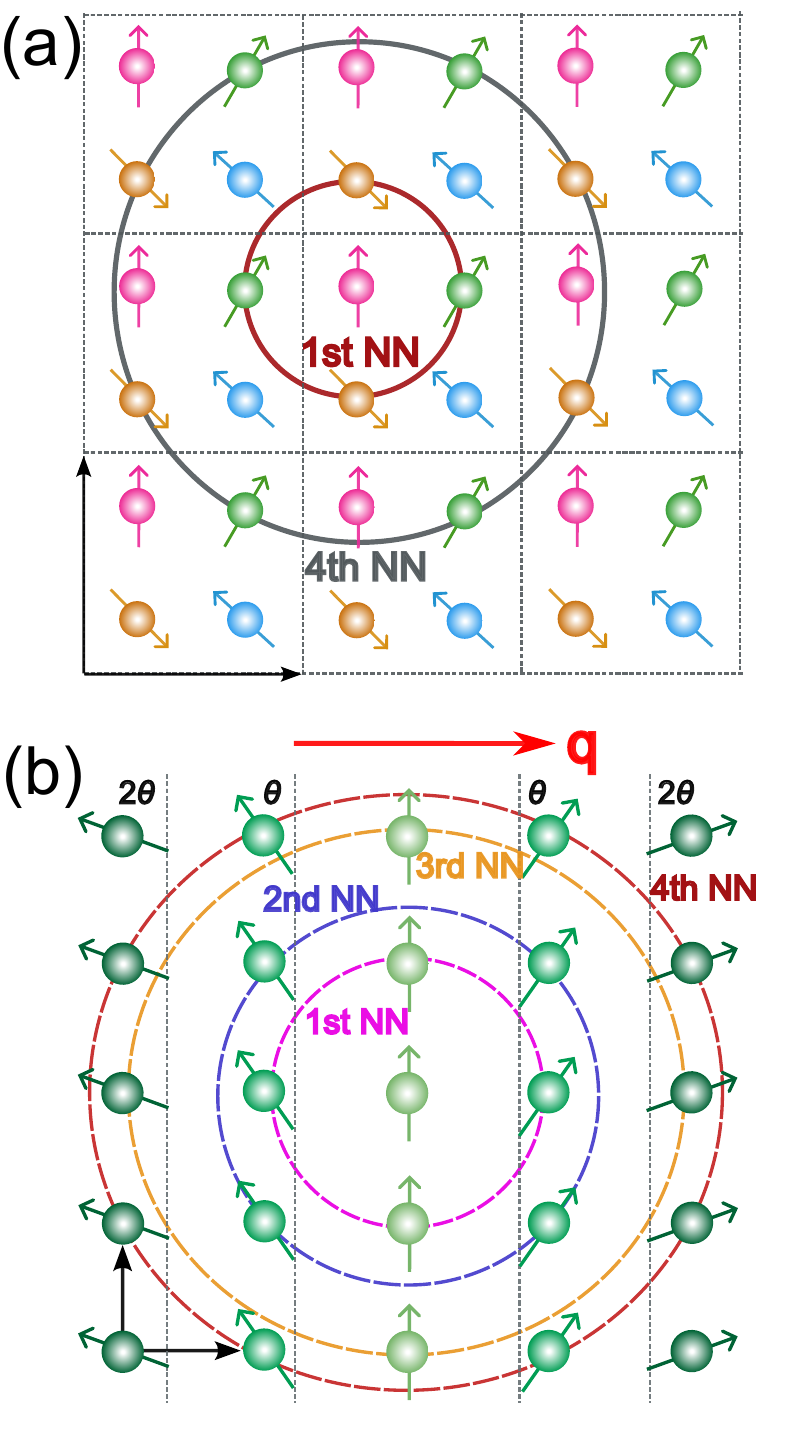}
	\caption{Schematic illustration of the linear dependence between neighbor shells for (a) a random spin state and (b) a spin spiral. (a) Magnetic random spin state in a two-dimensional lattice with four atoms per primitive cell. Different colors denote different atomic sublattices, and arrows indicate the corresponding magnetic moments. The structure shown is a $3 \times 3$ supercell. The first and fourth neighbor shells are labeled as red circles and grey circles. (b) Spin spiral in a two-dimensional lattice with one atom per primitive cell. The atomic magnetic moments are shown as arrows with different colors, reflecting their orientations in the spin-spiral state. The propagation vector of the spin spiral is indicated by the red arrow labeled $\mathbf{q}$. The first, second, third, and fourth neighbor shells are denoted by the pink, blue, orange, and brown circles, respectively. In all panels, the black arrows indicate the lattice vectors.}
	\label{figure:1}
\end{figure}

Here, we develop a neighbor-shell-based screening framework that analyzes linear dependence directly at the shell contributions for the magnetic configurations. Using only the structure, our method identifies which exchange neighbor shells can be independently resolved, without requiring magnetic configurations to be constructed or first-principles calculations to be performed. This enables the pre-screening of suitable magnetic configurations prior to DFT calculations, which reduces the computational cost and time required for exchange-parameter extraction, while extending the analysis from collinear states to noncollinear configurations.

To demonstrate how magnetic states affect the reliability of exchange interactions, we consider two representative classes of magnetic configurations: random spin states and spin spirals. The former offer a flexible and symmetry-unconstrained set of spin orientations, whereas the latter introduce a well-defined noncollinear modulation controlled by the propagation vector $\mathbf{q}$. In the random configurations, the linear dependence between neighbor shells can be illustrated using the two-dimensional lattice with four atoms per primitive cell as shown in Fig.~\ref{figure:1}(a). The example is a $3\times3$ supercell in which the four distinct sublattices are different colors and carry different magnetic moments. For a reference pink atom, the first neighbor shell contains two green and two orange sites, whereas the fourth neighbor shell contains four green and four orange sites. Under periodic boundary conditions, the interaction associated with the fourth neighbor shell ($S_4$) is always exactly twice that of the first-neighbor shell ($S_1$), i.e., $S_4 = 2S_1$, regardless of the choice of magnetic moments. Consequently, the fitting exchange parameters for $J_1$ and $J_4$ become linearly dependent for the random magnetic configurations constructed using this structure, making these parameters not uniquely identifiable during the fitting procedure.

Spin spirals constitute another commonly used type of magnetic configuration for extracting exchange interactions \cite{Sandratskii1991, Tung2011}. Despite their noncollinear nature, spin spirals can be computed within the primitive cell in the absence of SOC by using the GBT \cite{Kurz2004}. Nevertheless, they are also subject to the same issue of linear dependence among the different neighbor shells. This can be illustrated for spin-spiral configurations using the two-dimensional lattice with one atom per primitive cell shown in Fig.~\ref{figure:1}(b). For a given propagation vector $\mathbf{q}$, the magnetic moments rotate by fixed relative angles determined by the lattice translations. As a result, the shell interactions are constrained by the geometry of the spin spiral. In this example, the relation $S_4 = -2S_1 + 2S_2 + 2S_3$ holds, demonstrating that the fourth neighbor shell is linearly dependent on the first three shells. Therefore, the corresponding fourth-shell exchange parameter cannot be extracted independently. 

\subsection{Random spin states}
For a given structure, random spin states can be assigned within the primitive cell. Due to the periodic boundary conditions, this magnetic arrangement is then replicated throughout the crystal, thereby generating a complete magnetic configuration for the infinite lattice. For the Heisenberg model described by Eq.~\ref{eq:1}, by combining all symmetry-equivalent exchange interactions and retaining exchange couplings up to the $k$-th neighbor shell, we obtain

\begin{align}
	H_{\mathrm{eff}} = -\frac{1}{2} \frac{N}{N_{cell}}\sum_k J_k S_k,
	\label{eq:2}
\end{align}
where $J_k$ represents the exchange interaction for the $k$-th nearest neighbor, $N_{cell}$ is the number of magnetic sites in the unit cell, $N$ is the total number of magnetic sites as defined above, and $S_k$ denotes the total exchange interaction for the $k$-th neighbor shell, which can be calculated as

\begin{equation}
	S_k = \sum_{i}^{N_{cell}} \sum_{j}^{N_n} \mathbf{e}_i \cdot \mathbf{e}_j,
	\label{eq:3}
\end{equation}
where $N_{n}$ is the number of neighbor sites in the $k$-th neighbor shells. This shell contribution describes the total exchange interaction between each central atom and all atoms belonging to its $k$-th neighbor shell, and can be used to weight the exchange parameter $J_k$ in the fitting. The spin-correlation term $\mathbf{e}_i \cdot \mathbf{e}_j$ between the central atom $i$ and its neighboring atom $j$ is determined by the combinations of atoms within the unit cell due to the imposed boundary conditions, thereby making the corresponding shell contributions finite. We can define the exchange interaction pairs as,

\begin{equation}
	E_{ij} = \mathbf{e}_i \cdot \mathbf{e}_j \,\, (i,j=1,...,N_{cell}).
	\label{equation:2-2}
\end{equation}

By exploiting the spatial symmetry of the system, one finds that $E_{ij} = E_{ji}$, which implies that the self-interaction terms $E_{ii}$ are identical for all atoms in the cell. Since the magnetic moments are normalized, these diagonal terms satisfy $E_{ii}=1$, so that $E_{ii}$ contributes as a constant term. Physically, $E_{ii}$ represents the special periodic contribution associated with pairs related by lattice translations. Because this contribution is completely determined by translational symmetry, it remains unchanged for all random spin states constructed from this cell and merely shifts the total energy by the same constant amount. Therefore, it only provides a constant shift to the total energy and does not contribute to the energy differences between distinct spin arrangements. In contrast, for pairs of atoms that are not related by a lattice translation, the relative spin orientations may change from one magnetic configuration to another, and their exchange contributions vary accordingly. Therefore, the energy differences between distinct magnetic configurations originate from these configuration-dependent pair interactions, whereas the configuration-independent $E_{ii}$ terms contribute only a constant offset and do not affect the relative energies.

These configuration-dependent interactions terms constitute a basis set for describing the interactions within a given neighbor shell. The dimension of this basis can be calculated as $N_{b}=N_{cell}\,(N_{cell}-1)/2$.
Each $S_k$ can be decomposed into contributions from this finite-dimensional basis spanned by all configuration-dependent pairs $E_{ij}$,

\begin{equation}
	S_{k} = \sum_{i,j} C^k_{ij} E_{ij}.
	\label{eq:5}
\end{equation}
By computing the expansion coefficients $C^k_{ij}$ of $S_k$ in the basis for each $k$-th neighbor shells, the linear correlations between neighbor shells can be detected. The first neighbor shell that becomes linearly dependent on the preceding neighbor shell indicates the maximum neighbor-shell order for which the exchange parameters can be reliably determined for the given structure using random spin states. The contributions of neighbor shells beyond this cutoff cannot be uniquely separated from those of lower-order shells within the selected random-spin-state basis.

To extract exchange parameters beyond the maximum neighbor-shell order, one must consider either a unit cell with a different shape or a larger supercell. Changing the unit-cell shape does not alter the number of basis vectors in Eq.~\ref{eq:5}, but it may change their composition, thereby modifying the form of the neighbor-shell interactions and their linear dependence relations. On the other hand, enlarging the supercell size directly increases the number of basis vectors in Eq.~\ref{eq:5}, leading to more complex representation of the neighbor-shell interactions and potentially making previously correlated shells become linearly independent. Therefore, both methods can extend the range of exchange parameters that can be uniquely determined by different mechanisms.

\subsection{Spin spirals}
For spin spirals, however, the construction of a basis such as in the random-spin-state approach is generally impractical. In most cases, the spin-spiral period is not commensurate with the primitive unit cell, so that a direct real-space description requires a large supercell to accommodate the magnetic modulation \cite{Kurz2004}. This makes it computationally expensive to resolve neighbor-shell interactions. Therefore, a different basis is needed to represent the neighbor-shell interactions in spin spirals. A spin spiral is a noncollinear magnetic structure in which the magnetic moments rotate periodically along a propagation vector $\mathbf{q}$, characterized by the rotation axis and the cone angle $\beta$ between the magnetic moment and the rotation axis. Choosing the rotation axis as the $z$ direction, the normalized magnetic moment at site $i$ can be written as

\begin{equation}
    \mathbf{e}_i =
    \left(
    \sin\beta \cos\phi_i,\,
    \sin\beta \sin\phi_i,\,
    \cos\beta
    \right),
    \qquad
    \phi_i = \mathbf{q}\cdot\mathbf{R}_i+\xi^\alpha .
    \label{eq:6}
\end{equation}
Here, $\phi_i$ is the spin-rotation phase of site $i$, $\mathbf{R}_i$ is its lattice position, and $\xi^\alpha$ is an atom-dependent phase that accounts for the relative rotation of different magnetic atoms within the unit cell, with $\alpha$ labeling the magnetic atom in the unit cell. The total energy $E(\mathbf{q}, \beta, \xi^{\alpha})$ depends on the spin-spiral propagation vector, the cone angle, and the relative phases between the magnetic atoms in the unit cell. 

We first consider the simplest case of a flat spin spiral. In this configuration, all magnetic moments rotate in a plane perpendicular to a fixed rotation axis. For an arbitrary pair of atoms $i$ and $j$, their relative position vector can be written as $\mathbf{r}_{ij}=\mathbf{R} \pm \mathbf{r}^{\alpha}$, where \(\mathbf{R}\) is the lattice vector and \(\mathbf{r}^{\alpha}\) denotes the relative vector within the primitive cell. For a commensurate propagation vector, $\mathbf{q}$ can be written as $\mathbf{q} = q\cdot \mathbf{Q} = q\cdot (a, b, c)$, where $\mathbf{Q}=(a, b, c)$ is the primitive integer direction vector along the propagation vector of $\mathbf{q}$ and $q$ is the corresponding scaling factor. For example, $\mathbf{q}=(0.25,0.25,0.75)$ can be written as $0.25\cdot(1,1,3)$. For the flat spin spiral, the cone angle remains fixed at $\beta=90^{\circ}$ for each magnetic moment relative to the rotation axis, while the transverse component rotates periodically in the plane perpendicular to this axis, which is determined by the propagation vector $\mathbf{q}$ and additional atom-dependent phase $\xi^{\alpha}$ within the unit cell. The corresponding relative rotation angle between the central atom $i$ and its neighbor atom $j$ can then be expressed as

\begin{equation}
    \gamma = \mathbf{q} \cdot \mathbf{R} + \xi^{\alpha} = n\theta +\xi^{\alpha}.
    \label{eq:7}
\end{equation}

Where $n\in\mathbb{Z}$ denotes the number of times the rotation is applied, $\xi^{\alpha}$ is the additional phase introduced by $\mathbf{r}^{\alpha}$, and $\theta=2\pi q$ is the smallest rotation angle from cell to cell in a spin spiral configuration. For a given spin spiral with fixed $\mathbf{q}$ and $\xi^{\alpha}$, the relative angles between magnetic moments are not arbitrary, but take only discrete values set by $\theta$ and the additional atom-dependent phase $\xi^{\alpha}$, indicating that the relative orientation between magnetic moments is quantized by lattice periodicity. For spin spirals with the same propagation direction $\mathbf{Q}$ but different magnitudes $q$, although the minimum rotation angle between adjacent cells $\theta$ varies with $q$, the rotation time $n$ remains unchanged because it is determined only by the crystal symmetry and propagation direction $\mathbf{Q}$. As a result, the spin spiral configurations generated by different $q$ values along the same $\mathbf{Q}$ direction share the same geometric pattern, which leads to linear dependence among the exchange interactions of different neighbor shells for these spin spirals. The exchange interaction between the central atom $i$ and its neighbor atom $j$ can be calculated as

\begin{equation}
    \mathbf{e}_i \cdot \mathbf{e}_j = \cos \, (n\theta+ \xi^{\alpha}),
    \label{eq:8}
\end{equation}
and the exchange interaction for the $k$-th neighbor shell $S_k$ in the flat spin spiral can be defined as

\begin{equation}
	S_k^{\mathrm{flat}}  = \sum_{n,\alpha} C_{n\alpha} ^{k} \cos\,(n\theta+ \xi^{\alpha}).
	\label{eq:9}
\end{equation}

A more general situation is the conical spin spiral. In this case, each magnetic moment keeps a fixed cone angle $\beta$ with respect to the rotation axis, while its transverse component follows the same spiral rotation determined by the propagation vector $\mathbf{q}$. Therefore, the phase relation between the central atom $i$ and its neighboring atom $j$ is unchanged from the flat-spin-spiral case and remains $\phi_{ij}=n\theta+\xi^\alpha$. The exchange interaction between the two magnetic moments can now be written as
\begin{equation}
\mathbf{e}_i\cdot\mathbf{e}_j = \sin^2\beta \cos(n\theta+\xi^\alpha) + \cos^2\beta.
\label{eq:10}
\end{equation}
Accordingly, the contribution from the $k$-th neighbor shell in the conical spin spirals can be written as
\begin{equation}
S_k^{\mathrm{cone}} = N_k\cos^2\beta + \sin^2\beta \, S_k^{\mathrm{flat}},
\label{equation:conical-2}
\end{equation}
where $N_k$ denotes the coordination number of the $k$-th neighbor shell, i.e., the number of equivalent exchange paths associated with this shell. The first term in Eq.~\ref{equation:conical-2} represents the shell contribution from the cone angle of the spin spiral. For a fixed cone angle $\beta$, this contribution to the energy of spin spiral is constant and independent of the spin-spiral propagation vector $\mathbf{q}$. This configuration-independent contribution has no effect on the energy difference between distinct conical spin spirals with the identical cone angles. By contrast, the second term represents the exchange contribution from the rotating transverse components, which contains the same $\mathbf{q}$-dependent phase information as the corresponding flat spin spiral, with its amplitude rescaled by $\sin^2\beta$. Thus, for a fixed cone angle, a conical spin spiral introduces no new spin-rotation pattern compared with the flat spin spiral. As a result, the configuration-dependent part of $S_k^{\mathrm{cone}}$ is governed by the same basis as $S_k^{\mathrm{flat}}$, so the linear dependence of different neighbor shells remains unchanged.

The shared configuration-dependent $\cos{(n\theta+\xi^{\alpha})}$ terms in the flat and conical spin spirals provide a natural basis for representing the configuration-dependent contribution of each neighbor shell. This approach is analogous to the one used for random spin states to describe the interactions within the $k$-th neighbor shell. Therefore, for spin spirals with identical propagation direction $\mathbf{Q}$, this provides a convenient framework to analyze the linear independence among different neighbor-shell interactions, with coefficients fully determined by the crystal geometry and propagation direction. When such spin spirals are used to fit the Heisenberg Hamiltonian, the number of reliable fitting parameters is limited. As in the analysis of random spin states, the first neighbor shell that becomes linearly dependent on the former shells identifies the maximum neighbor-shell order for which the exchange parameters can be reliably determined for the given structure using these spin spirals.

A special class of exchange paths occurs when the neighboring atoms are related to the central atom by a lattice vector $\mathbf{R}$ and lie in the plane normal to the propagation direction $\mathbf{Q}$ passing through the central atom, which corresponds to $n=0$ and $\xi^{\alpha}=0$. Consequently, the relative rotation angle between such atoms and the central atom is always zero for all spin spirals along this propagation direction, so that the corresponding interactions are constant. Their contribution to the energy is therefore constant for all spin-spiral states considered along this direction. Such terms are configuration-independent, which means that they only add a constant offset to the total energy and do not affect the energy differences between distinct spin spirals. 

\subsection{Workflow for magnetic-configuration design}
Based on the neighbor-shell linear-dependence analyses developed above for two representative classes of magnetic configurations (random spin states and spin spirals), we now formulate a workflow for magnetic-configuration design. The key idea is to treat a magnetic configuration not only as a magnetic state computed by first-principles calculations, but also as a source of symmetry- and periodicity-constrained neighbor-shell contributions to the total energy. By analyzing the linear independence of these shell contributions before performing DFT calculations, one can evaluate whether a given set of candidate magnetic configurations contains sufficient information to extract exchange parameters up to a desired neighbor cutoff.

\begin{figure}[htbp]
	\centering
	\includegraphics[width=0.75\linewidth]{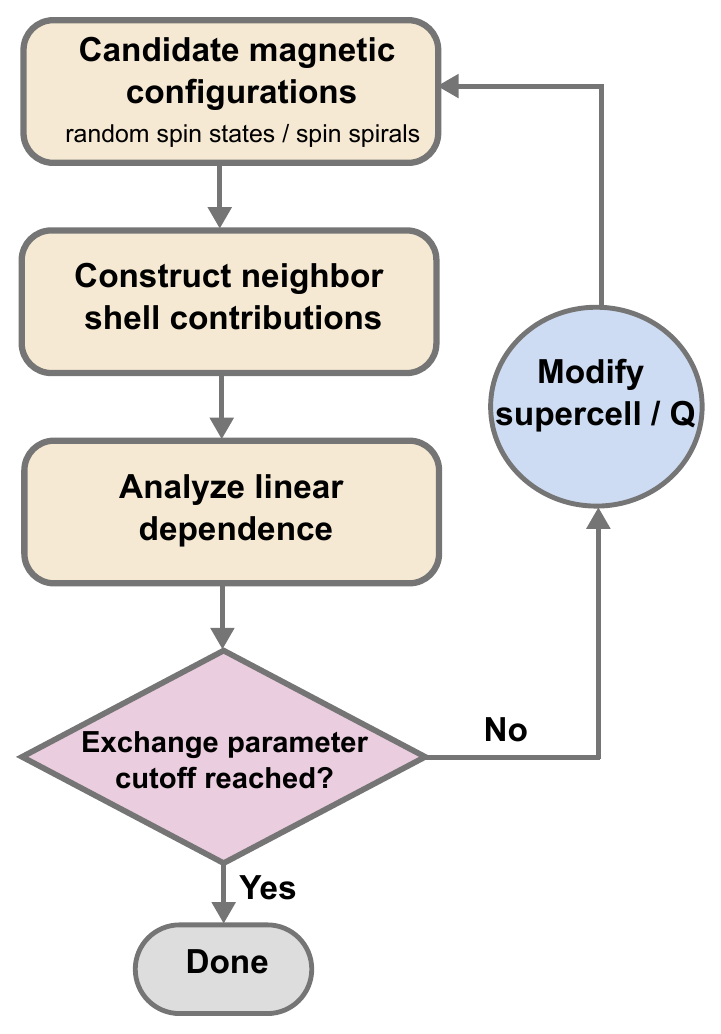}
    \caption{Workflow for magnetic-configuration design.}
	\label{figure:2}
\end{figure}

Figure~\ref{figure:2} outlines the proposed workflow for the magnetic-configuration design, where random spin states and spin spirals are considered as two representative examples of candidate magnetic configurations for exchange-parameter extraction. In practice, the linear-dependence analysis was implemented as a \textsc{python} script. For a given crystal structure and magnetic-configuration family, the script first constructs the neighbor shells and collects all exchange interactions of magnetic pairs belonging to each shell into a single shell contribution. Each exchange path is then mapped onto the corresponding configuration-dependent basis, from which the shell-contribution coefficient matrix is obtained. As discussed above, the configuration-dependent bases are the independent spin-pair interactions allowed by the chosen supercell for random spin states, and the phase-dependent terms $\cos(n\theta+\xi^\alpha)$ for spin spirals.

The resulting coefficient matrix describes how each neighbor shell contributes to the magnetic energy within the selected configuration family. Its rank is analyzed by adding neighbor shells sequentially, starting from the shortest-range interaction. A shell is considered linearly independent if its inclusion increases the rank, indicating that it provides new magnetic information for the total energy. If the rank remains unchanged, the shell contribution can be represented by the preceding shells, and the corresponding exchange parameters cannot be uniquely resolved. The first dependent shell defines the resolvable neighbor-shell cutoff, beyond which the exchange parameters cannot be reliably determined from the selected magnetic configurations.

A challenge of magnetic-configuration design is that the information carried by the configurations depends strongly on how they are constructed. For random magnetic states, the accessible neighbor-shell information is determined by the size and shape of the chosen supercell. For spin spirals, it is determined by the propagation direction, which controls how different neighbor shells are combined. Therefore, an arbitrary choice of supercell or spin-spiral direction may not provide enough independent information for the targeted number of exchange parameters. To overcome this limitation, we use an iterative screening procedure. Starting from an initial configuration family, either random spin states or spin spirals, we evaluate the linear independence of the neighbor-shell contributions and identify the maximum number of exchange parameters that can be reliably resolved. If this number is lower than the desired neighbor-shell cutoff, the configuration family is updated by modifying the supercell for random spin states or by changing the propagation direction for spin spirals. The procedure is repeated until the selected configurations provide sufficient independent information for the targeted exchange-parameter extraction.

\subsection{One-atom model system: bcc Fe}
After establishing the workflow for magnetic-configuration design, we apply it to bcc Fe as a representative system for assessing random spin states and spin spirals in exchange-parameter extraction. Since bcc Fe contains only one magnetic atom in the primitive cell, the neighbor-shell contributions are not complicated by additional sublattice-dependent phase relations. 

\subsubsection{Random spin states}
\begin{figure}[htbp]
	\centering
	\includegraphics[width=0.85\linewidth]{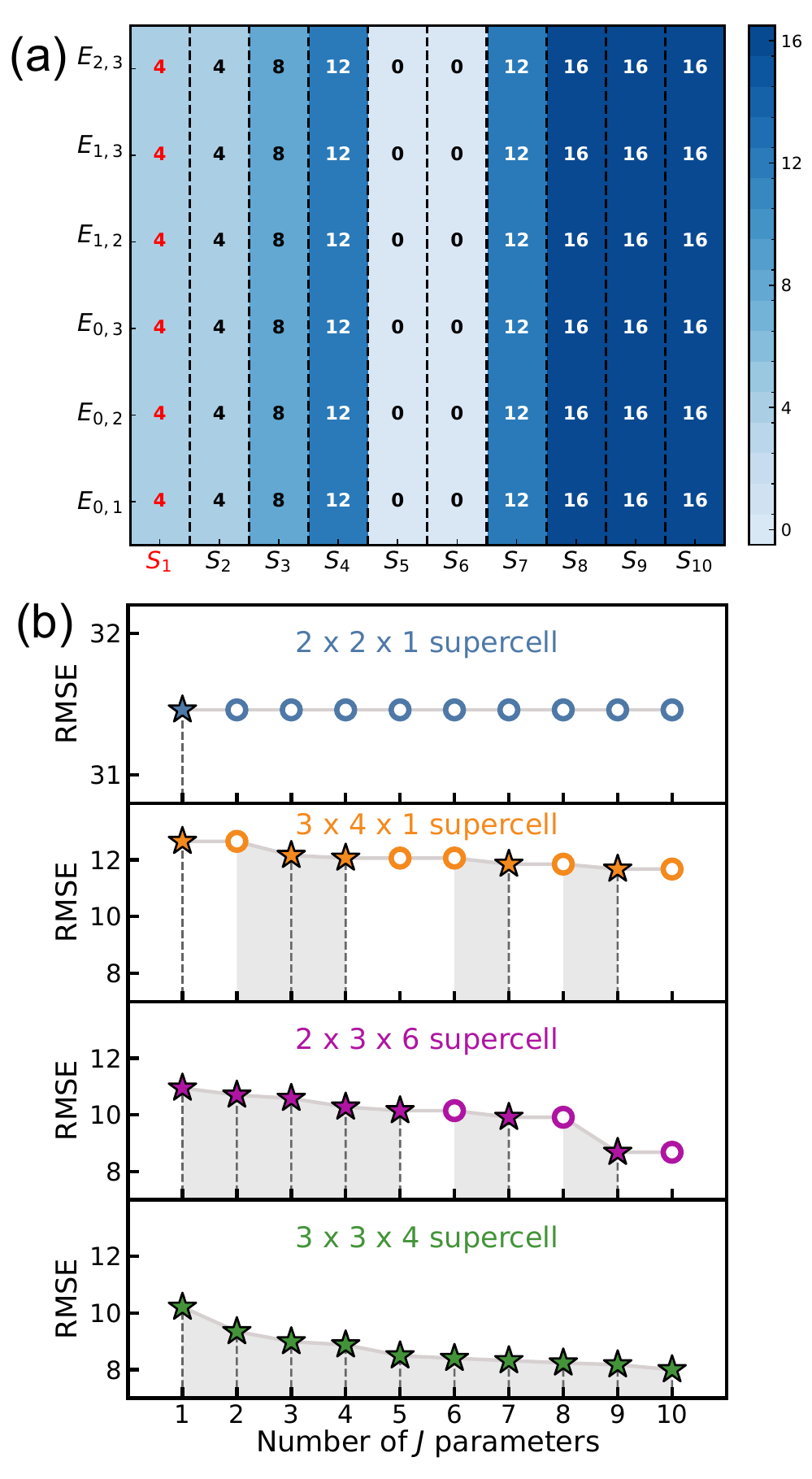}
    \caption{Linear dependence analysis of neighbor-shell interactions in bcc Fe for random spin states. Panel (a) shows the allowed interaction basis functions for the first to tenth neighbor shells. The $x$ axis labels the neighbor shells $S_n$, while the $y$ axis denotes the pair-interaction basis functions $E_{ij}$. The numbers in each cell indicate the corresponding expansion coefficient for a given shell in this basis. For example, for the second neighbor shell, $S_2 =4\left(E_{0,1}+E_{0,2}+E_{0,3}+E_{1,2}+E_{1,3}+E_{2,3}\right).$ The hatched columns with red labels denote linearly independent neighbor-shell contributions. Panel (b) shows the fitting RMSE as a function of the number of fitted exchange parameters $J$ for different supercell sizes. Stars and circles denote linearly independent and dependent shell contributions, respectively. The dashed lines and gray shaded regions mark RMSE-changing points.}
	\label{figure:3}
\end{figure}

For random spin states, the available shell information is determined by the spin-pair correlations in the unit cell. A $2\times2\times1$ Fe supercell is constructed from the primitive cell to generate random magnetic configurations, containing four magnetic Fe sites in the primitive cell, labeled as 0, 1, 2, and 3, whose spin orientations can be varied independently to generate random magnetic states. For any magnetic atom $j$, its magnetic moment is taken to be the same as that of the corresponding atom $k$ folded back into the chosen periodic supercell due to the boundary conditions. That is, if $\mathbf{r}_j = \mathbf{r}_k + n\mathbf{R}$, where $k$ denotes the index of the atom in the chosen supercell corresponding to atom $j$, then $\mathbf{e}_j = \mathbf{e}_k$. According to Eq.~\ref{eq:5}, the configuration-dependent exchange interaction terms $\mathbf{e}_i \cdot \mathbf{e}_j$ between a central atom $i$ and its neighboring atom $j$ define a six-dimensional basis for representing the interactions within each neighbor shell. The exchange interaction associated with each neighbor shell can therefore be expanded in terms of these configuration-dependent bases as

\begin{equation}
    S_k = \sum_{\substack{i,j=0 \\ i\ne j}}^{3} C^{k}_{ij} \, E_{ij},
    \label{eq:12}
\end{equation}
where $S_k$ denotes the total interaction contribution from the $k$-th neighbor shell, and $C^{k}_{ij}$ are the corresponding expansion coefficients. For this $2\times2\times1$ supercell, the expansion coefficients shown in Fig.~\ref{figure:3}(a) indicate that only $S_1$, corresponding to the first-neighbor exchange shell, is linearly independent. By contrast, $S_2$ does not introduce any new independent information, since it can be written as a linear combination of $S_1$, i.e., $S_2=S_1$. Consequently, the highest neighbor-shell order whose exchange parameter can be uniquely extracted from these random magnetic configurations is limited to the first shell, so that only $J_1$ can be reliably determined. It is worth noting that the expansion coefficients of the fifth- and sixth-neighbor shells, $S_5$ and $S_6$, vanish identically in the present basis. This arises because the atoms belonging to these shells are related to the central atom through lattice translation vectors, so their relative spin orientations are fixed by translational symmetry and cannot vary with the magnetic arrangement inside the unit cell. These interactions therefore contribute only a configuration-independent constant term and have no effect on the energy differences among distinct random magnetic states. Overall, the expansion coefficients show that the $2\times2\times1$ supercell does not contain sufficient linearly independent information to uniquely determine any exchange parameter beyond $J_1$.

This model prediction is further confirmed by directly fitting the exchange parameters to the DFT-computed spin-spiral energies, as shown in Fig.~\ref{figure:3}(b). Here, the fitting quality is evaluated using the root-mean-square error (RMSE) for different numbers of fitted exchange parameters, where a smaller RMSE indicates a better agreement between the fitted model and the DFT energies. As the number of fitting parameters increases from 1 to 2, the RMSE remains unchanged, demonstrating that inclusion of an additional exchange parameter does not improve the description of the energies for random spin states. This behavior is fully consistent with the prediction of our linear independence analysis, which indicates that $J_1$ is the only exchange parameter that can be uniquely resolved in this case.

While using random magnetic configurations to introduce more reliable fitting parameters, a larger supercell or a different unit-cell shape is required. Here, a larger $3\times4\times1$ supercell containing twelve atoms with random magnetic spins in the chosen supercell is constructed. In this case, the number of configuration-dependent pair-interaction basis functions increases to 66, reflecting a greater complexity of the neighbor-shell interactions compared with the previous case. The expansion coefficients of each neighbor-shell interaction in this basis can be obtained from Eq.~\ref{eq:5} (see Supplementary Materials). The model analysis reveals that the first, third, fourth, seventh, and ninth neighbor shells are linearly independent. Since the first appearance of linear dependence sets the upper boundary on the neighbor-shell order for which the exchange parameters can be robustly determined, the highest exchange parameter that can be reliably extracted is still $J_1$. The fitting results obtained for the random spin states in the $3\times4\times1$ supercell are in agreement with the prediction of our method, showing that the RMSE changes only for fits including one, three, four, seven, and nine parameters, as shown in Fig.~\ref{figure:3}(b). Although the $3\times4\times1$ supercell introduces more configuration-dependent pair-interaction basis functions and leads to a more complex pattern of shell-level linear independence compared to $2\times2\times1$, it does not increase the number of exchange parameters that can be robustly extracted, because the second neighbor shell is still interrupted by the first shell.

To overcome this limitation, we next consider a larger $2\times3\times6$ supercell containing 36 magnetic Fe atoms with independently assigned random spin directions and the number of configuration-dependent pair-interaction basis functions increases to 630. Applying the same shell-resolved linear-independence analysis to this supercell, we find that the coefficients associated with the first, second, third, fourth, fifth, seventh, and ninth neighbor shells are linearly independent (see Supplementary Material). Therefore, the maximum number of exchange parameters that can be reliably extracted is increased to five due to the first occurrence of linear dependence in the sixth neighbor shell. This prediction is further supported by the fitting results obtained from DFT calculations, where the change in fitting performance follows the shell-level independence identified by the model analysis, as shown in Fig.~\ref{figure:3}(b).

The above comparison shows that increasing the supercell size can improve the number of reliably extractable exchange parameters, but this improvement comes at the higher cost of DFT calculation. An alternative strategy is to change the supercell shape while keeping the total number of magnetic atoms fixed. We consider a $3\times3\times4$ supercell, which also contains 36 magnetic Fe atoms and therefore has the same dimension of configuration-dependent pair-interaction basis functions as the $2\times3\times6$ supercell. The model analysis shows that the first ten neighbor shells are all linearly independent (See Supplementary Materials), indicating that up to ten exchange parameters can be reliably extracted. This prediction is again confirmed by the DFT fitting results as shown in Fig.~\ref{figure:3}(b): as the number of fitted exchange parameters increases from one to ten, the RMSE changes continuously, consistent with the absence of linear dependence within the first ten shells. Compared with the $2\times3\times6$ supercell, the $3\times3\times4$ supercell increases the number of reliably extractable exchange parameters from five to ten without increasing the number of atoms. Although the dimension of the pair-interaction basis remains the same, changing the supercell shape modifies how different neighbor-shell interactions are projected onto this basis, and therefore changes their linear-dependence relations. Thus, under the same computational cost, choosing an appropriate supercell shape provides a more efficient route to resolving a larger number of reliable exchange interactions.

\subsubsection{Spin spirals}

After analyzing how random spin states encode neighbor-shell interactions through real-space spin-pair correlations, we now turn to the second representative class of magnetic configurations: spin spirals. In this case, the shell contributions are governed by the phase factors associated with the propagation direction $\mathbf{Q}$. Following the argument of Ref.~[\citen{Daglum2026}], we focus on conical spin spirals in bcc Fe with a fixed cone angle of $\beta=30^\circ$, while varying the magnitude of $q$ along a given propagation direction $\mathbf{Q}$. Since bcc Fe contains only one magnetic atom in the primitive cell, no additional sublattice phase factor is present, namely $\xi^{\alpha}=0$. The pair interaction between magnetic moments can therefore be expressed in terms of configuration-dependent basis $\cos(n\theta)$, as given in Eq.~\ref{equation:conical-2}. Accordingly, the contribution from the $k$-th neighbor shell can be written as

\begin{equation}
	S_k^{\mathrm{cone}} = N_k\cos^2\beta + \sin^2\beta\sum_{n} C_{n} ^{k} \cos\,(n\theta),
	\label{equation:3-2-1}
\end{equation}
where $n \in \mathbb{Z}\setminus\{0\}$ and $N_k$ is the coordination number of the $k$-th neighbor shell as mentioned above. Since $N_k\cos^2\beta$ is independent of the magnetic configuration and has no contribution to the energy differences between spin spirals, the linear dependence is determined by the expansion coefficients $C_n^k$ associated with the configuration-dependent basis $\cos(n\theta)$.

\begin{figure*}[htbp]
	\centering
	\includegraphics[width=0.7\linewidth]{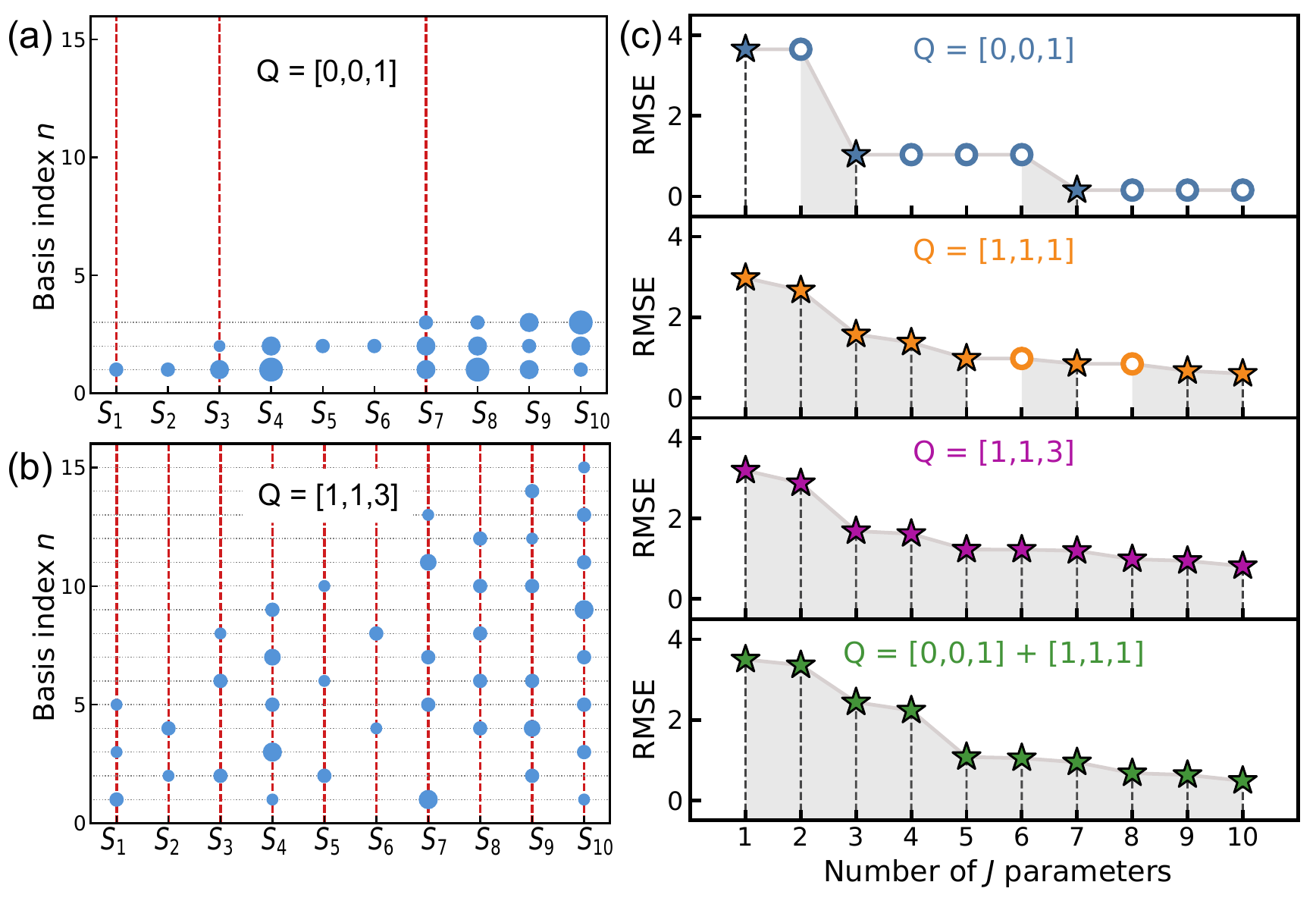}
	\caption{Linear dependence model analysis of neighbor-shell interactions in bcc Fe for spin spirals along (a) $\mathbf{Q} = (0,0,1)$ and (b) $\mathbf{Q} = (1,1,3)$. The $x$ axis labels the neighbor shells, and the $y$ axis denotes the basis index $n$. The circle size indicates the magnitude of the corresponding expansion coefficient. The red vertical dashed lines indicate the neighbor shells that introduce linearly independent contributions. Panel (c) shows the fitting RMSE as a function of the number of fitted exchange parameters $J$ for different directions. Stars and circles denote linearly independent and dependent shell contributions, respectively. The dashed lines and gray shaded regions mark RMSE-changing points.}
	\label{figure:4}
\end{figure*}

The first example propagation direction for analyzing the first ten neighbor shells is $\mathbf{Q}=(0,0,1)$, where the propagation vector can be written as $\mathbf{q}=q(0,0,1)$. The expansion coefficients $C_{n} ^{k}$ of the first ten neighbor shells are shown in Fig.~\ref{figure:4}(a). From this analysis, $S_1$, $S_3$, and $S_7$ corresponding to the exchange interactions in the first, third, and seventh neighbor shells, are predicted to be linearly independent. However, $S_2$ can be expressed as a linear combination of the first neighbor shell, indicating that there is no additional independent information in the second neighbor shell. As a result, extending the interaction range beyond the first neighbor shell does not increase the number of reliably extracted exchange parameters. Although $S_3$ and $S_7$ are linearly independent of the preceding shells, the linear dependencies among $S_2$ and $S_1$ indicate that the number of exchange parameters that can be robustly determined remains limited to one. This is further confirmed by the fitting quality measured by the RMSE for the DFT-calculated spin-spiral energy dispersion along $\mathbf{Q}=(0,0,1)$ for different numbers of fitting parameters, as shown in Fig.~\ref{figure:4}(c). The RMSE changes when the number of fitting parameters is increased from 0 to 1, from 2 to 3, and from 6 to 7, indicating that the first, third, and seventh neighbor shells provide linearly independent contributions. However, no change in the RMSE is observed when the number of fitting parameters is increased from 1 to 2, which shows that the second-neighbor shell is already linearly dependent on the preceding ones. Due to the first appearance of linear dependence setting the limit for the reliable extraction of exchange parameters, only $J_1$ can be robustly determined in this case, which is in full agreement with the model prediction.

For another example propagation direction, $\mathbf{Q}=(1,1,1)$, the spin spiral provides a different projection of the basis vectors and therefore gives access to more complex information on the exchange shells. According to the model analysis (See Supplementary Materials), eight out of the first ten neighbor shells are linearly independent for this propagation direction. However, the first linear dependence already appears at the sixth neighbor shell, meaning that the information associated with $S_6$ cannot be independently resolved from the preceding shells. Consequently, $J_6$ cannot be reliably determined from the fitting, and the number of exchange parameters that can be robustly extracted is limited by this first occurrence of linear dependence to five. This conclusion is also supported by the RMSE behavior of the DFT spin-spiral energy fitting: when the number of fitting parameters is increased from 5 to 6, the fitting quality remains unchanged, indicating that the sixth neighbor shells does not provide an independent contribution to the fit.

The above two examples propagation directions contain useful independent information for a limited number of exchange parameters, but they are not sufficient when a larger number of parameters, for example up to ten neighbor shells, needs to be reliably fitted. To overcome this limitation in spin spiral configurations, two possible strategies can be considered. The first one is to search for a suitable low-symmetry propagation direction. $\mathbf{Q} = (1,1,3)$ is one of such less-symmetric directions. For spin spirals in bcc Fe along $\mathbf{Q} = (1,1,3)$, the same model can be applied to analyze the linear relations among neighbor shells, as shown in Fig.~\ref{figure:4}(b). The model analysis shows that the first ten neighbor shells are all linearly independent along this direction, which is further supported by the fitting results of DFT-calculated spin spiral energy dispersion: the fitting quality changes continuously as the number of fitting parameters is increased from 1 to 10, indicating that each additional neighbor shell provides new information, as show in Fig.~\ref{figure:4}(c). Thus, the first ten neighbor shells are confirmed to be linearly independent, in agreement with our model analysis. This indicates that, for exchange interactions up to the tenth-neighbor shell, $\mathbf{Q}=(1,1,3)$ should be selected instead of the $\mathbf{Q}=(0,0,1)$ and $\mathbf{Q}=(1,1,1)$ directions, as it provides sufficient independent information for the reliable extraction of Heisenberg exchange parameters.

The second method is to combine spin-spiral dispersions along multiple propagation directions. Here, we combine the $\mathbf{Q}=(0,0,1)$ and $\mathbf{Q}=(1,1,1)$ directions. In the single-direction analyses, the sixth and eighth neighbor shells remain linearly dependent and cannot be independently resolved. However, our model analysis shows that the first ten neighbor shells become linearly independent when the two directions are considered simultaneously. This means that the combined spin spirals restore the missing structural information of these shells and allow at least ten reliable exchange parameters to be extracted. The DFT fitting results confirm this prediction, since the RMSE decreases consistently up to the tenth fitting parameter, as shown in Fig.~\ref{figure:4}(c). Although effective, the multi-direction approach is more computationally demanding than using a single well-chosen low-symmetry direction, as it requires additional expensive DFT spin-spiral calculations. Therefore, in practical applications, it is preferable to first search for a suitable low-symmetry propagation direction that can distinguish the desired neighbor shells within a single spin-spiral path.

\subsubsection{Exchange parameters transferability}

The reliability of the extracted exchange parameters depends on both the linear independence of the neighbor-shell contributions and the inclusion of significant noncollinear magnetic information in the configurations. When all targeted neighbor shells are linearly independent, each exchange parameter contributes distinct magnetic information to the total energy and can be uniquely determined. In this case, the extracted parameters are expected to be transferable and can be used to predict the energies of other magnetic configurations. In this section, we examine this transferability using bcc Fe as a test case. Since spin-spiral calculations can be performed within the primitive cell using the GBT while still incorporating non-collinear magnetic information, they provide a good balance between computational efficiency and the accuracy of the exchange-parameter fitting. We therefore focus on the exchange parameters extracted from spin-spiral calculations.

\begin{figure}[htbp]
	\centering
	\includegraphics[width=0.8\linewidth]{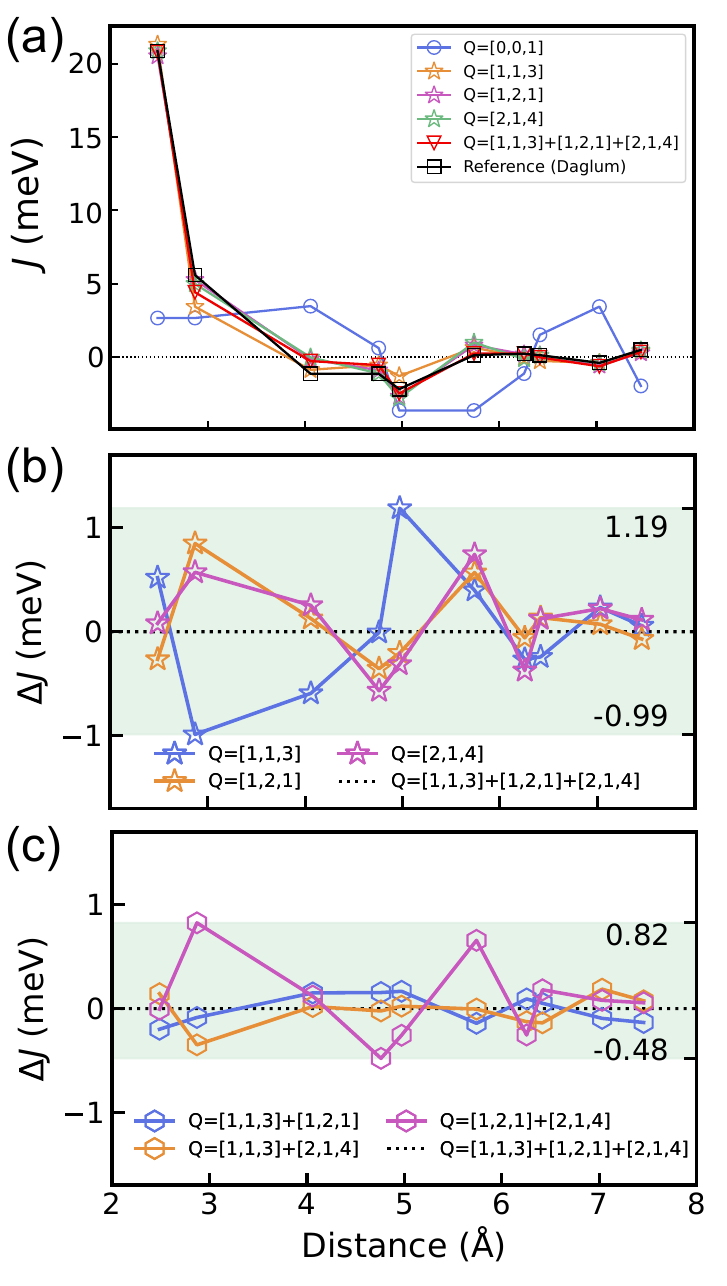}
	\caption{(a) Exchange parameters $J$ obtained from different fitting schemes, compared with previous study: Daglum $et$ $al$ \cite{Daglum2026}. Different colors and markers distinguish the fitting schemes, while the black squares denote the reference values. (b,c) Differences $\Delta J$ relative to the parameters obtained by simultaneously fitting the three linearly independent directions. (b) single-direction fits and (c) combined fits using two directions. The colored curves and markers identify the corresponding fitting directions or direction combinations. The green shaded region indicates the maximum variation range of the first ten exchange parameters.}
	\label{figure:5}
\end{figure}

To assess the quality of the fitted exchange parameters, we compare the first ten $J$ parameters obtained using different spin-spiral fitting schemes, as shown in Fig.~\ref{figure:5}(a). The exchange parameters obtained from the three propagation directions, $\mathbf{Q}=(1,1,3)$, $\mathbf{Q}=(1,2,1)$, and $\mathbf{Q}=(2,1,4)$, for which the contributions from the first ten neighbor shells are linearly independent (See Supplementary Materials), are in close agreement with those reported by Daglum \textit{et al.}~\cite{Daglum2026}. In that work, the exchange parameters were derived by applying the inverse Fourier transform to self-consistent spin-spiral energies calculated on a uniform $10\times10\times10$ $\mathbf{q}$-grid. Moreover, the close agreement among these three linearly independent propagation directions further shows that the extracted exchange parameters are not sensitive to the particular propagation direction, but instead converge toward a consistent set of exchange parameters. However, for the $\mathbf{Q}=(0,0,1)$ direction in which only three of the first ten neighbor-shell contributions are linearly independent, the resulting parameters exhibit a different behavior, with $J_1$, $J_2$, and $J_3$ remaining at comparable magnitudes, indicating that exchange parameters extracted from a linearly dependent propagation direction are less reliable. 

As additional linearly independent propagation directions are incorporated into the fitting, the dataset samples the magnetic interactions over a broader region of reciprocal space and provides more independent information on the contributions from different neighbor shells. This increased information improves the robustness of the extracted exchange parameters and reduces their dependence on the selection of spin spirals. As shown in Fig.~\ref{figure:5}(b) and (c), the deviations from the parameters obtained by fitting all three linearly independent directions become smaller when the number of fitting directions is increased from one to two, and the variation range of the difference of extracted exchange parameters is reduced, indicating that the exchange parameters become converged as additional independent directions are included in the fitting. Therefore, to obtain the reliable exchange parameters in practical calculations, one should select appropriate linearly independent propagation direction or combine multiple propagation directions in the fitting.

\begin{figure*}[htbp]
	\centering
	\includegraphics[width=0.8\linewidth]{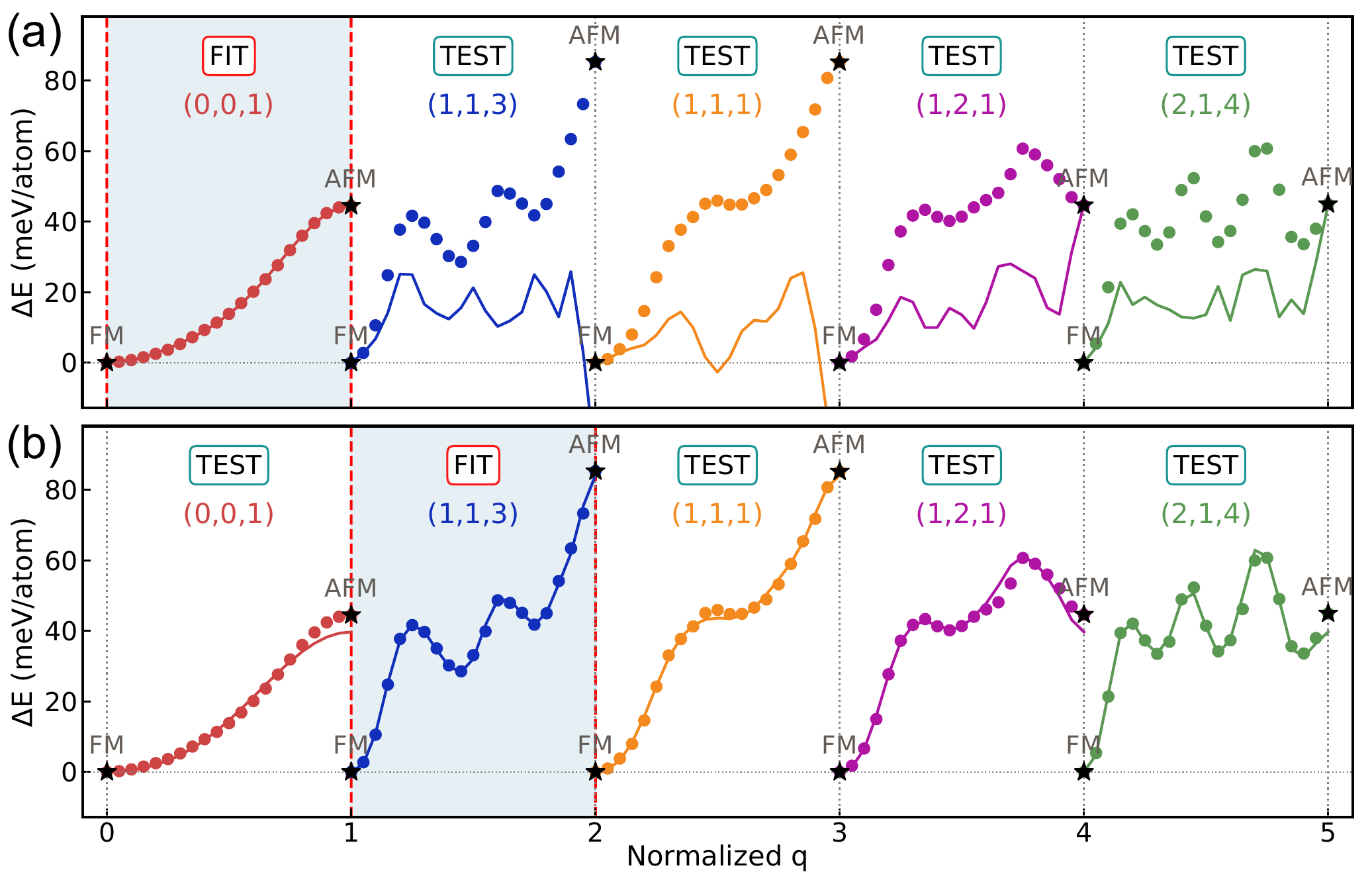}
	\caption{The spin spiral energy dispersions of bcc Fe calculated using the ten exchange parameters fitted in Fig.~\ref{figure:5} from the DFT results along $\mathbf{Q}=(0,0,1)$ in (a) and $\mathbf{Q}=(1,1,3)$ in (b). The $y$ axis shows the energy difference $\Delta E$ with respect to the ferromagnetic (FM) state. The $x$ axis represents the normalized propagation vector $\mathbf{q}$ along different $\mathbf{Q}$ directions; for each direction, the path is normalized to the interval from 0 to 1. Points denote the DFT-calculated energies, dashed lines denote the fitted energies, and star symbols indicate the FM and antiferromagnetic (AFM) states. Data with the same color belong to the same $\mathbf{Q}$ direction, as indicated by the labels at the top of each panel. The blue-shaded FIT region indicates the fitting direction, while the unshaded TEST regions indicate the prediction directions.}
	\label{figure:6}
\end{figure*}

Figure~\ref{figure:6} compares the fitting results of bcc Fe obtained using ten exchange parameters $J$ for the DFT spin-spiral energies along two example propagation directions, $\mathbf{Q}=(0,0,1)$ and $(1,1,3)$. The fitted parameters are further used to predict the spin-spiral energy dispersions along other directions in order to assess their quality. For $\mathbf{Q}=(0,0,1)$, although the fit to the training data is excellent, the predictions for other directions are very poor. By contrast, for $\mathbf{Q}=(1,1,3)$, the fit is not only excellent for the training data, but also provides accurate predictions for the spin spiral energy dispersions along other directions. This difference originates from the linear dependence among neighbor shells. Along the $\mathbf{Q}=(1,1,3)$ direction, the first ten neighbor shells are all linearly independent from above analysis and therefore provide sufficient information for determining the exchange interactions. However, along the $\mathbf{Q}=(0,0,1)$ direction, linear dependencies already appear from the second neighbor shell onward, so the corresponding spin-spiral energies do not contain enough independent information to determine robust exchange parameters.

\subsection{Two-atom model system: MnF$_2$}
To examine how the proposed shell-level analysis extends from a single-sublattice system to a multi-sublattice magnetic material, we further consider $\mathrm{MnF_2}$ as a complementary test case. For random spin states, the analysis is mainly determined by the spin-pair correlations generated within the chosen supercell, and the same procedure can be directly applied to other materials. Therefore, we focus here on spin spirals, for which the presence of more than one magnetic atom in the primitive cell introduces additional sublattice-dependent phase factors. $\mathrm{MnF_2}$ provides a representative example of this situation, as its primitive cell contains two magnetic Mn atoms. Here, we consider flat spin spirals where the cone angle $\beta=90^{\circ}$ In this case, the linear dependence of neighbor-shell contributions is not governed solely by the propagation direction $\mathbf{Q}$ and the interatomic translation vectors, but also affected by the relative phases associated with the magnetic atoms inside the primitive cell.

Due to the presence of two magnetic atoms in the primitive cell of MnF$_2$, the spin-spiral description requires an additional phase factor $\xi^{\alpha}$. The relative positions associated with the magnetic atoms in the primitive cell are given by $\mathbf{r}^{\alpha}=0$, $\mathbf{R}^{\prime}$, and $-\mathbf{R}^{\prime}$, where $\mathbf{R}^{\prime}=\frac{1}{2}\mathbf{a}_1+\frac{1}{2}\mathbf{a}_2+\frac{1}{2}\mathbf{a}_3$, and $\mathbf{a}_1$, $\mathbf{a}_2$, and $\mathbf{a}_3$ are the primitive lattice vectors. For a homogeneous spin spiral, all magnetic atoms including those within the same primitive cell, acquire a phase determined by the propagation vector $\mathbf{q}$, and the additional phase factor can be calculated as $\xi^{\alpha}=\mathbf{q}\cdot\mathbf{r}^{\alpha}$. The corresponding relative rotation angle $\gamma$ between the central atom $i$ and its neighbor atom $j$ can be expressed from Eq.~\ref{eq:7} as

\begin{equation}
    \gamma = (n + \xi)\theta,
    \label{eq:14}
\end{equation}
where $\theta=2\pi q$ represents the smallest rotation angle between neighboring cells in the spin spiral, while $n \in \mathbb{Z}$ and $\xi=0,\pm\frac{1}{2}$. As discussed above in the last section, special cases arise when $n=0$ and $\xi=0$, for which the corresponding relative rotation angles are configuration-independent. Such terms merely give a constant contribution to the total energy and therefore do not affect the energy differences among different spin spirals. The remaining configuration-dependent contributions provide a description of the interaction for the $k$th neighbor shell, which can be obtained from Eq.~\ref{eq:9} as

\begin{equation}
	S_k^{\mathrm{flat}} = \sum_{n,\xi} C_{n\xi} ^{k} \cos\,[(n+\xi)\theta].
	\label{eq:15}
\end{equation}

\begin{figure}[htbp]
	\centering
	\includegraphics[width=1\linewidth]{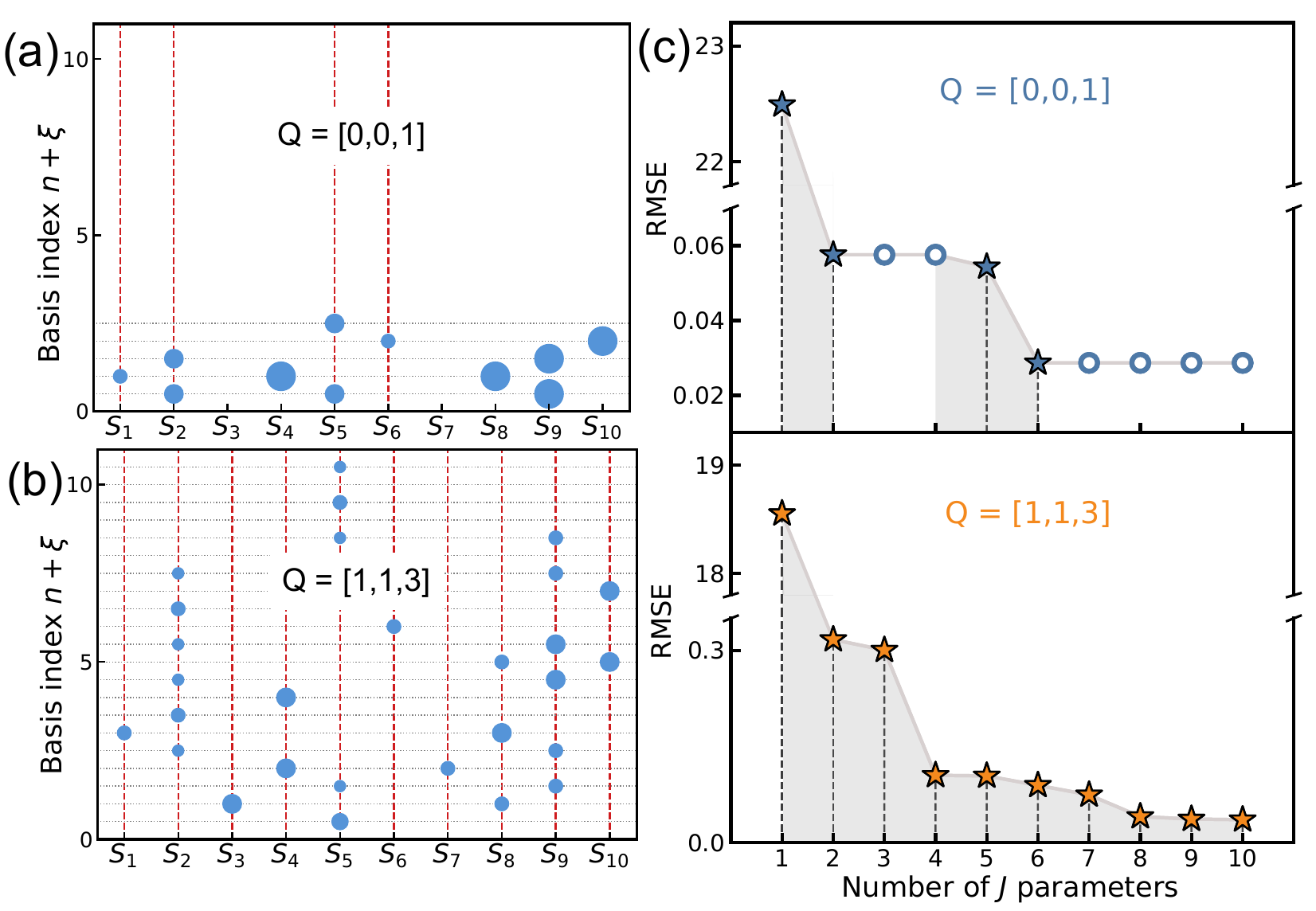}
	\caption{Linear dependence analysis of neighbor-shell interactions in $\mathrm{MnF_2}$ for spin spirals along (a) $\mathbf{Q} = (0,0,1)$ and (b) $\mathbf{Q} = (1,1,3)$. The $x$ axis labels the neighbor shells, and the $y$ axis denotes the basis index $n+\xi$. The circle size indicates the magnitude of the corresponding expansion coefficient. The red vertical dashed lines indicate the neighbor shells that introduce linearly independent contributions. Panel (c) shows the fitting RMSE as a function of the number of fitted exchange parameters $J$ for $\mathbf{Q} = (0,0,1)$ (blue) and $\mathbf{Q} = (1,1,3)$ (orange). Stars and circles denote linearly independent and dependent shell contributions, respectively. The dashed lines and gray shaded regions mark RMSE-changing points.}
	\label{figure:7}
\end{figure}

These coefficients $C_{n\xi}^k$ provide a convenient framework for analyzing the dependence of shell interactions on the propagation direction $\mathbf{Q}$. We next consider two representative propagation directions to illustrate how the additional sublattice phase affects the linear dependence of neighbor-shell contributions. The first direction is chosen as $\mathbf{Q}=(0,0,1)$, and the expansion coefficients for the first ten neighbor shells are shown in Fig.~\ref{figure:7}(a). While the first, second, fifth, and sixth shells are formally linearly independent, the third and fourth shells are linearly dependent on the preceding shells. Once including the third and fourth shell, the extracted exchange parameters lose their robustness and can no longer be considered reliable. This remains true even if the fifth and sixth shells which contain additional linearly independent components are further included. In practice, this limits the reliable extraction of exchange parameters from the spin-spiral energy dispersion of $\mathrm{MnF_2}$ along $\mathbf{Q}=(0,0,1)$ to the first two neighbor shells. It should be noted that the expansion coefficients associated with the third and seventh neighbor shells, $S_3$ and $S_7$, vanish identically in the present basis as shown in Fig.~\ref{figure:7}(a). This originates from the fact that the atoms in these shells are related to the central atom by lattice translation vectors and simultaneously lie in the plane containing the central atom and perpendicular to the propagation direction. As a consequence, their relative spin orientations are fixed by the crystal structure for a given propagation direction. They therefore do not carry any configuration-dependent information relevant to the energy difference between different spin spirals, and the corresponding expansion coefficients vanish in the present expansion. The corresponding fitting qualities to the DFT spin-spiral energy dispersions obtained using different numbers of exchange parameters are displayed in Fig.~\ref{figure:7}(b). The RMSE is found to change only when the number of fitted parameters reaches one, two, five, and six, while in all other cases it remains identical to that of the preceding results. This behavior confirms that only four linearly independent combinations are present within the first ten neighbor shells, but the maximum number of reliable exchange parameters is up to 2, which is consistent with the model analysis.

A similar analysis can be carried out \(\mathrm{MnF_2}\) along another example direction \(\mathbf{Q}=(1,1,3)\). The results in Fig.~\ref{figure:7}(c) show that the first ten neighbor shells are all linearly independent in this case. This conclusion is consistent with the corresponding fits to the DFT spin-spiral energy dispersion, for which the fitting quality improves continuously as the number of exchange parameters is increased from 1 to 10, as shown in Fig.~\ref{figure:7}(c). Such behavior confirms that the first ten neighbor-shell contributions are linearly independent. Therefore, for this material as well, the $\mathbf{Q}=(1,1,3)$ spin-spiral direction is appropriate for extracting exchange interactions up to the tenth-neighbor shell, enabling ten reliable exchange parameters to be determined.

The fitting results for $\mathrm{MnF_2}$ (see Supplementary material) lead to the same conclusion as in bcc Fe. The spin-spiral energy dispersions along the $\mathbf{Q}=(0,0,1)$ and $\mathbf{Q}=(1,1,3)$ directions were fitted using ten exchange parameters and then tested by predicting energy dispersions along other propagation directions. Although both fittings can reproduce their corresponding training data, their transferability is very different. The parameters obtained from $\mathbf{Q}=(1,1,3)$ give a good prediction of the other spin-spiral directions, whereas those fitted from $\mathbf{Q}=(0,0,1)$ show poor transferability. This difference is consistent with the model analysis: the first ten neighbor shells are linearly independent for $\mathbf{Q}=(1,1,3)$, whereas for $\mathbf{Q}=(0,0,1)$, linear dependence already appears at the third-neighbor shell. Therefore, as in bcc Fe, a small fitting error for the training data alone is not sufficient to guarantee reliable exchange parameters; the underlying neighbor-shell contributions must also be linearly independent.

\section{Discussion}\label{sec3}

In this work, we have introduced an efficient strategy for selecting magnetic configurations for the extraction of Heisenberg exchange interactions by analyzing the linear independence of neighbor-shell contributions associated with a chosen family of magnetic configurations. For a given target number of exchange parameters, the analysis determines whether the selected configurations contain sufficient independent information to resolve the corresponding neighbor shells. Instead of generating many candidate configurations and assessing them only after first-principles calculation, the proposed criterion provides an a priori estimate of the neighbor-shell information carried by each configuration family, reducing the trial-and-error effort usually involved in selecting magnetic configurations and the computational cost of first-principles calculations. We demonstrated this idea using two representative classes of magnetic configurations: random spin states and spin spirals. For random spin states, the resolvability of exchange interactions is controlled by the spin-pair correlations generated within a finite supercell, so both the supercell size and shape play an important role in determining which neighbor shells can be independently extracted. For spin spirals, the shell contributions are instead governed by the phase factors associated with the propagation vector $\mathbf{Q}$ and, in multi-sublattice systems, by the relative phases between magnetic atoms in the primitive cell. 

We test the proposed analysis on bcc Fe using both random spin states and spin spirals, and further test it on a multi-sublattice system $\mathrm{MnF_2}$ using spin spirals. The maximum number of reliably extractable parameters predicted by the shell-level linear-independence analysis is consistent with the changes observed in direct fitting to DFT energies, confirming that the method captures how supercell size and shape in random states, as well as the propagation direction $\mathbf{Q}$ and additional sublattice-dependent phase in spin spirals, control the independent resolvability of neighbor-shell contributions. The transferability tests further demonstrate that only exchange parameters extracted from linearly independent neighbor-shell contributions can be reliably transferred to other magnetic configurations, showing that linear independence is a necessary condition for robust exchange-parameter extraction. These test cases show that the reliability of fitted exchange parameters is not determined only by the accuracy of the calculated total energies, but also by how the selected magnetic configurations encode independent information about different exchange neighbor shells.

In practice, this provides a guideline for exchange-parameter fitting: adjust the cell shape or supercell size for random spin states, and select or combine suitable propagation directions for spin spirals, so that all targeted neighbor-shell contributions are linearly independent. The present analysis provides these example magnetic configurations, while more general magnetic textures may introduce additional degrees of freedom. For example, for non-homogeneous spin spirals or more complex spin textures, the relevant shell contributions may depend on both the propagation direction and the spin-pair relations within the unit cell, especially in multi-sublattice systems. Therefore, combining the proposed shell-level analysis with magnetic-moment stability checks and transferability validation will be important for extending the method to more complex magnetic Hamiltonians and materials. More generally, the concept of using linear dependence of neighbor shells for evaluation of magnetic textures for parameter extraction is not restricted to exchange interactions or Heisenberg-like models. It might also be relevant for other fitting schemes in which model parameters are expanded in terms of neighbor-shell contributions. Thus, we believe that our approach is a noteworthy step forward in increasing the quality and feasibility of parameter extraction in magnetism and potentially beyond.

\section{Method}\label{sec4}
\subsection{Computational details}
All density-functional theory (DFT) calculations were performed using the Vienna Ab initio Simulation Package (VASP) \cite{Kresse1993,Kresse1994,Kresse1996} and the projector augmented-wave (PAW) method \cite{Blochl1994}. The exchange-correlation energy was treated using the local-density approximation (LDA) \cite{Perdew1992}. The constrained local-moment approach was employed to constrain the directions of the atomic magnetic moments in the magnetic configurations \cite{Ma2015}. Carefully converged calculations were carried out to obtain reliable total energies and magnetic moments. The plane-wave kinetic-energy cutoff was set to $600$~eV, and the electronic self-consistency convergence threshold was set to $10^{-5}$~eV. The optimized k-point grids for each test case were $20\times20\times20$ (Fe) and $9\times9\times13$ (MnF$_2$). For all calculations, the numerical parameters were chosen to ensure that the relative energies between different magnetic configurations were well converged.

\section{Data availability}\label{sec5}
All data used in this manuscript are available on the \href{https://doi.org/10.24435/materialscloud:zg-1w}{https://doi.org/10.24435/materialscloud:zg-1w}.

\section{Code availability}\label{sec6}
The underlying code is available on \href{https://github.com/modl-uclouvain}{https://github.com/modl-uclouvain}.

\vspace{15mm}
\bibliographystyle{sn-nature}
\bibliography{ref}

\section{Acknowledgements}\label{sec7}
B.L. acknowledges the funding from the China Scholarship Council (Grant No. 202406100014). S.v.M. is a Postdoctoral Researcher of the Fonds de la Recherche Scientifique - FNRS. G.-M.R. is a Research Director of the Fonds de la Recherche Scientifique - FNRS. Computational resources have been provided by the Consortium des \'Equipements de Calcul Intensif (C\'ECI), funded by the Fonds de la Recherche Scientifique de Belgique (F.R.S.-FNRS) under Grant No. 2.5020.11 and by the Walloon Region. The present research also benefited from computational resources made available on Lucia, the Tier-1 supercomputer of the Walloon Region, infrastructure funded by the Walloon Region under the grant agreement n$^{\circ}$1910247.

\section{Author Contributions}\label{sec8}
The project was conceived by B.L., S.v.M., and G.-M.R. B.L. developed the code, performed the DFT calculations, and drafted the manuscript. S.v.M. and G.-M.R. supervised the work and revised the manuscript. All authors contributed to the analysis of the results and the writing of the manuscript.

\section{Competing interests}\label{sec9}
The authors declare no competing interests.

\section{Additional information}\label{sec10}
This publication is accompanied by supplementary material.

\end{document}